\documentclass[journal,twoside,web]{ieeecolor}
\usepackage{generic}
\usepackage{cite}
\usepackage{amsmath,amssymb,amsfonts}
\usepackage{algorithmic}
\usepackage{graphicx}
\usepackage{textcomp}
\usepackage[outdir=./]{epstopdf}

\usepackage{hyperref}

\def\BibTeX{{\rm B\kern-.05em{\sc i\kern-.025em b}\kern-.08em
    T\kern-.1667em\lower.7ex\hbox{E}\kern-.125emX}}
\begin{document}
\title{FPGA-Based In-Vivo Calcium Image Decoding for Closed-Loop Feedback Applications}
\author{Zhe Chen, \IEEEmembership{Member, IEEE}, Garrett J. Blair, Chengdi Cao, Jim Zhou, Daniel Aharoni, Peyman Golshani, Hugh T. Blair, and Jason Cong, \IEEEmembership{Fellow, IEEE}
\thanks{This work was supported by NSF NeuroNex DBI-1707408, and NIH MH122800. The authors would like to thank Dr. Changliang Guo and Jeffrey Johnson for their support on the miniscope device and the Ethernet interface development.}
\thanks{Zhe Chen and Jim Zhou were with the Computer Science Department, UCLA, Los Angeles, CA 90095 (email: marchzhe@gmail.com). Chengdi Cao and Jason Cong are with the Computer Science Department, UCLA (email: cong@cs.ucla.edu).}
\thanks{Hugh T. Blair is with the Psychology Department, UCLA (email: tadblair@ucla.edu). Daniel Aharoni and Peyman Golshani are with the Department of Neurology, UCLA. Garrett J. Blair was with Psychology Department, UCLA, and he is with the NYU Center for Neural Science, New York, NY 10003, USA.}}

\maketitle

\begin{abstract}
Miniaturized calcium imaging is an emerging neural recording technique that has been widely used for monitoring neural activity on a large scale at a specific brain region of rats or mice. Most existing calcium-image analysis pipelines operate offline. This results in long processing latency, making it difficult to realize closed-loop feedback stimulation for brain research. In recent work, we have proposed an FPGA-based real-time calcium image processing pipeline for closed-loop feedback applications. It can perform real-time calcium image motion correction, enhancement, fast trace extraction, and real-time decoding from extracted traces. Here, we extend this work by proposing a variety of neural network based methods for real-time decoding and evaluate the tradeoff among these decoding methods and accelerator designs. We introduced the implementation of the neural network based decoders on the FPGA, and showed their speedup against the implementation on the ARM processor. Our FPGA implementation enables the real-time calcium image decoding with sub-ms processing latency for closed-loop feedback applications.

\end{abstract}


\section{Introduction} \label{sec:1}

Calcium imaging is an emerging method that has been widely used for observing neural activity at a large scale in neuroscientific research. The miniaturized calcium imaging microscope is a device that can be head mounted on a live mouse or a rat for monitoring firing activity from hundreds of neurons at single cell resolution \textit{in vivo}, while allowing the rodent to perform behavioral tasks freely in the lab environment \cite{Ghosh2011}. UCLA leads the efforts in developing a series of open-sourced miniaturized calcium imaging microscopes ("miniscopes" \cite{Aharoni2019}). These miniscopes can record calcium activity from rats' or mice's brains and transfer them over meter-long flexible coaxial cable to a remote data acquisition (DAQ) board \cite{Guo2021}. They have gained popularity quickly among neuroscientific research labs and have accelerated brain research for many subfields related to memory, behavior and disease, to name a few.

Several complete calcium image analysis pipelines have been proposed to extract neuron activity from the raw calcium image recordings. \cite{Giovannucci2019} proposed a Python-based pipeline that has been widely used for one-photon calcium image analysis. It consists of separate motion correction \cite{Pnevmatikakis2017} and signal extraction \cite{Zhou2018} steps. The signal extraction relies on the constrained nonnegative matrix factorization (CNMF) approach to identify spatial cell footprints and temporal calcium traces simultaneously. \cite{Lu2018} provided a Matlab-based calcium image analysis pipeline, which was also based on the CNMF approach. It employs a long short-term memory (LSTM) inference approach to refine neuron activity detected by a Gaussian mixture model. These pipelines have been widely used in offline calcium image analysis. However, it is hard to be deployed online, because the CNMF approach requires a full stack of images as input. Though some recent work \cite{Friedrich2021} demonstrated that the CNMF-based pipeline achieved real-time throughput on general-purpose CPUs, it is still challenging to implement this on the embedded hardware with short processing latency for closed-loop feedback applications.

Some other research works \cite{Li2019, Rubin2019, Liu2021, Lee2019} attempt to decode behavioral information from the calcium images. \cite{Li2019} used a k-Means method to predict forelimb reach directions based on averaged calcium images. \cite{Rubin2019} decoded mice's behavioral states on a linear track by utilizing Laplacian Eigenmaps to extract internal representation of behaviors from high-dimensional neural activity patterns. Though these pioneering works show promising behavioral decoding results, they require batch processing, so they are difficult to be implemented for the online decoding with short latency. \cite{Wang2019} combines principal component analysis (PCA) with the linear support vector machine (SVM) to perform decoding during a pressing movement based on offline extracted calcium traces \cite{Lu2018}. Some recent works \cite{Liu2021, Lee2019} proposed SVM based methods for online calcium image decoding, however, they are evaluated on general purpose CPUs and there is still a lack of embedded hardware and electronic interface for demonstration of short processing latency for closed-loop feedback capability.

\begin{figure}
\centerline{\includegraphics[width=3.5in]{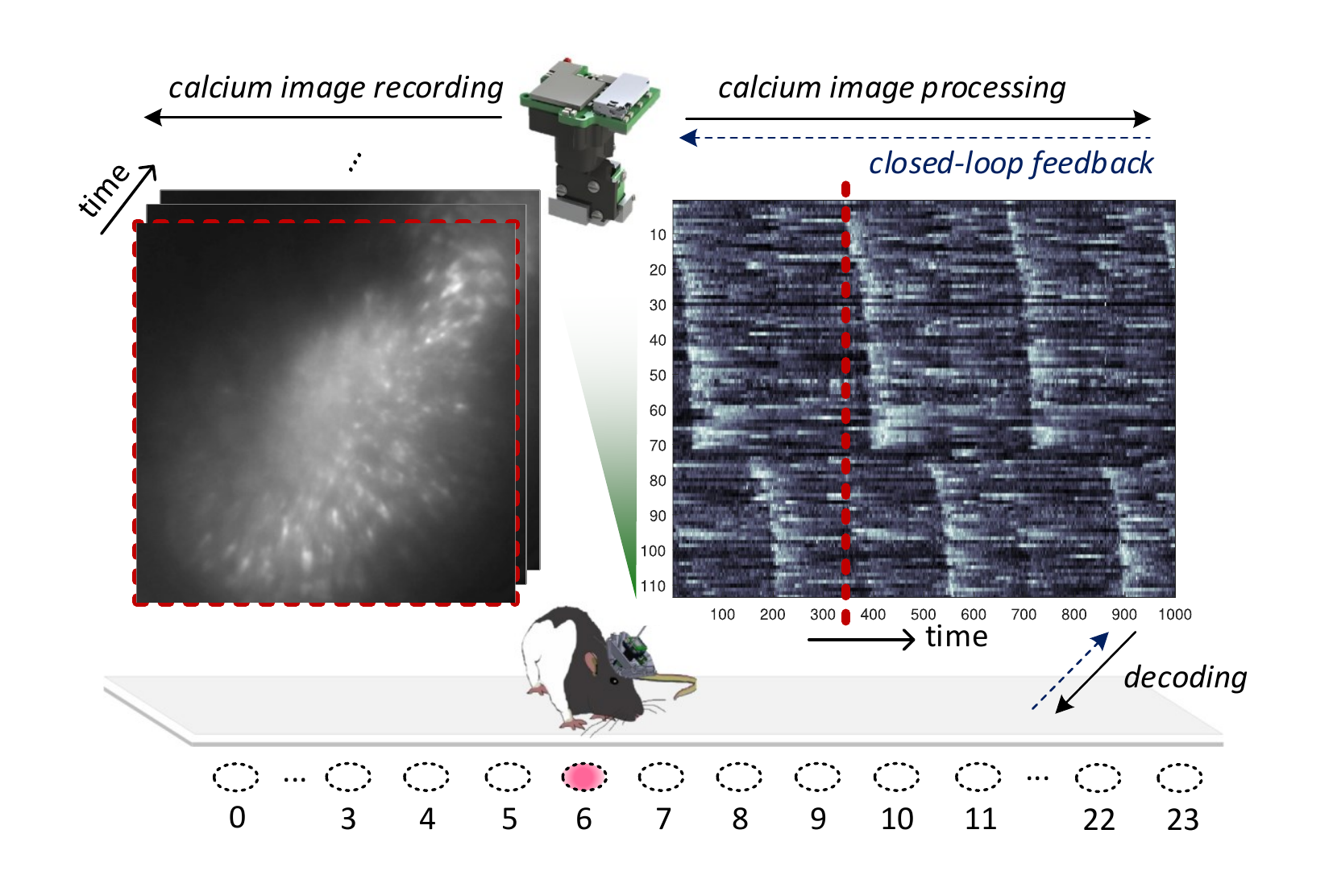}}
\caption{Closed-loop feedback capability enabled by real-time calcium image processing and decoding for miniscope devices: As the rat runs back and forth on a linear track, the computation kernel performs calcium image processing and decoding at each frame.}
\label{fig:1}
\end{figure}

Fig. \ref{fig:1} illustrates the main target this paper aims to achieve --- the closed-loop feedback capability enabled by real-time calcium image processing and decoding for miniscopes. Such closed-loop feedback capability brings three strict requirements: 1) The computation requires short and deterministic processing latency, in order to generate rapid and precise feedback based on event detection and decoding results from the calcium images. Recent advancement in voltage imaging \cite{Kazemipour2019} enables kilohertz frame rate, which escalates the requirements for short processing latency 2) The computation kernel must have low hardware cost and energy consumption, especially considering the integration of the computation kernel into the miniscope device. 3) The computation hardware has to be reconfigurable, as it may require adjustments of parameters involved in the calcium image processing across days during the experiments.

In prior work, we have proposed an FPGA-based real-time calcium image processing and decoding pipeline \cite{Chen2019, Chen2020, Chen2021, Chen2022_ISCAS, Chen2022_eLife}. Here we report several advances and improvements to our real-time calcium image processing system, including improvements to the trace extraction accelerator and the introduction of a variety of neural network based calcium image decoding methods. In addition, we introduced the implementation of the neural network based decoders on the FPGA. We showed that the FPGA implementation can achieve significant speedup over the implementation on the embedded ARM processor and enable real-time calcium image decoding with sub-ms latency.

The paper is arranged as follows: In Section \ref{sec:2}, we provide a brief review of our previously proposed real-time calcium image processing pipeline. In Section \ref{sec:3}, we present design and optimization details of the trace extraction accelerator. In Section \ref{sec:4}, we propose different neural network based methods and accelerator designs for the calcium image decoding, and evaluate the tradeoff among these decoding accelerators. Section \ref{sec:5} presents the implementation of the FPGA-based closed-loop calcium image processing pipeline. Section \ref{sec:6} summarizes the performance of this work and discusses related works in the field. Section \ref{sec:7} draws the conclusion.

\section{Review of Real-Time Processing Pipeline} \label{sec:2}

\begin{figure}
\centerline{\includegraphics[width=3.3in]{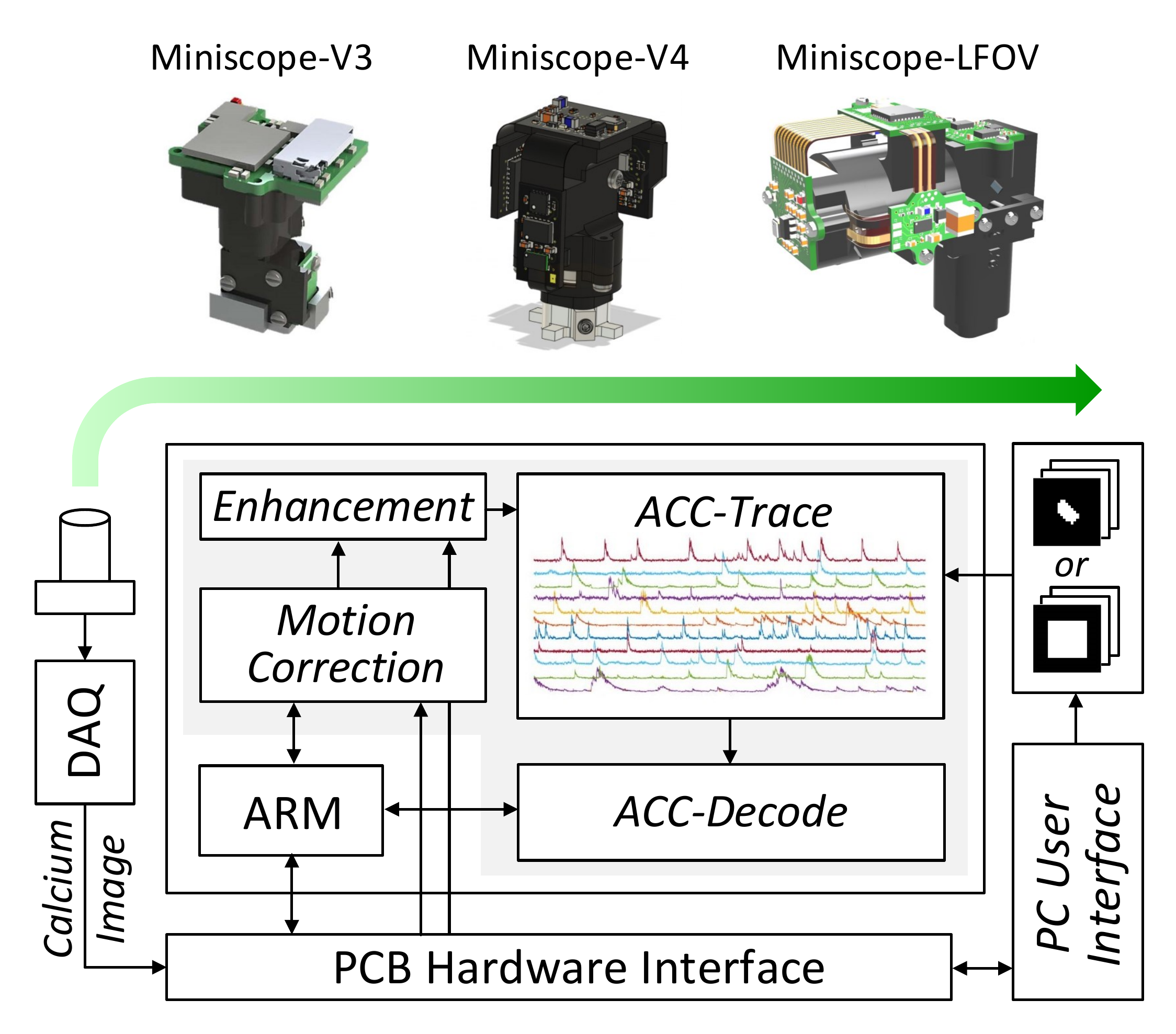}}
\caption{Review of DeCalciOn: The first FPGA-based real-time calcium image processing pipeline \cite{Chen2022_eLife}.}
\label{fig:2}
\end{figure}

\begin{figure}
\centerline{\includegraphics[width=3.5in]{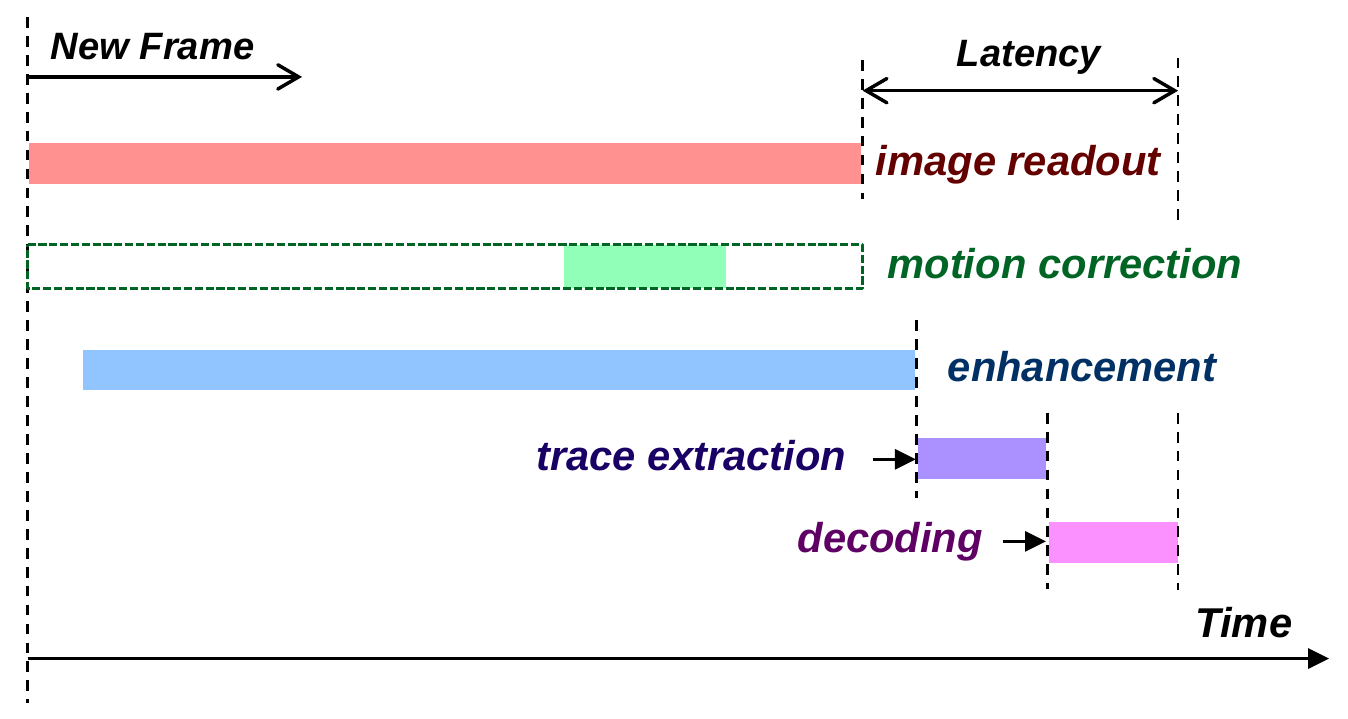}}
\caption{Timing diagram for the calcium image processing pipeline.}
\label{fig:3}
\end{figure}

In the past years, we proposed and developed the DeCalciOn – an FPGA-based real-time calcium image processing pipeline for closed-loop feedback applications \cite{Chen2022_eLife}. Fig. \ref{fig:2} shows the overall architecture of the processing pipeline. It includes customized accelerators for the motion correction, the image enhancement, the trace extraction, and the decoding. Fig. \ref{fig:3} shows the timing diagram of the DeCalciOn pipeline. The proposed accelerators operate in a streaming fashion and do not require off-chip memory access throughout the processing stages. As the complete pipeline and the accelerators for the motion correction and image enhancement have been introduced in \cite{Chen2022_eLife, Chen2019, Chen2020}, in this work we mainly focus on the implementation detail and the optimization for the trace extraction accelerator ACC-Trace and the decoding accelerator ACC-Decode.

\section{ACC-Trace: Fast Trace Extraction} \label{sec:3}

\textcolor{black}{In this session, we further elaborate the trace extraction accelerator: ACC-Trace, and discuss the optimization on fast trace extraction for both the cell-based and tile-based contours.}

\subsection{ACC-Trace Accelerator} \label{sec:3.1}

\textcolor{black}{Our proposed ACC-Trace accelerator has the 1-D systolic array architecture, as Fig. \ref{fig:4} shows. It consists of a chain of $J$} tracing elements (TE). Each TE contains the trace extraction logic and a local contour buffer, and it takes in and shifts an 8-bit pixel vi with its corresponding 9-bit row and column indices (${r}_{i}$,${c}_{i}$) per clock cycle. The local buffer can store up to $K$ pieces of cell contour information for the trace extraction. For a cell Cell($j$,$k$) mapped to the $k$th piece of the $j$th TE, we store not only an ${{N}_{C}}\times{{N}_{C}}$ binary mask ${Q}_{j,k}$, but also the cell center location (${R}_{j,k}$,${C}_{j,k}$). Suppose the calcium image has $N$ pixels, the 16-bit trace value ${f}_{jk}$ for the Cell($j$,$k$) can be derived by:

\begin{figure}
\centerline{\includegraphics[width=3.5in]{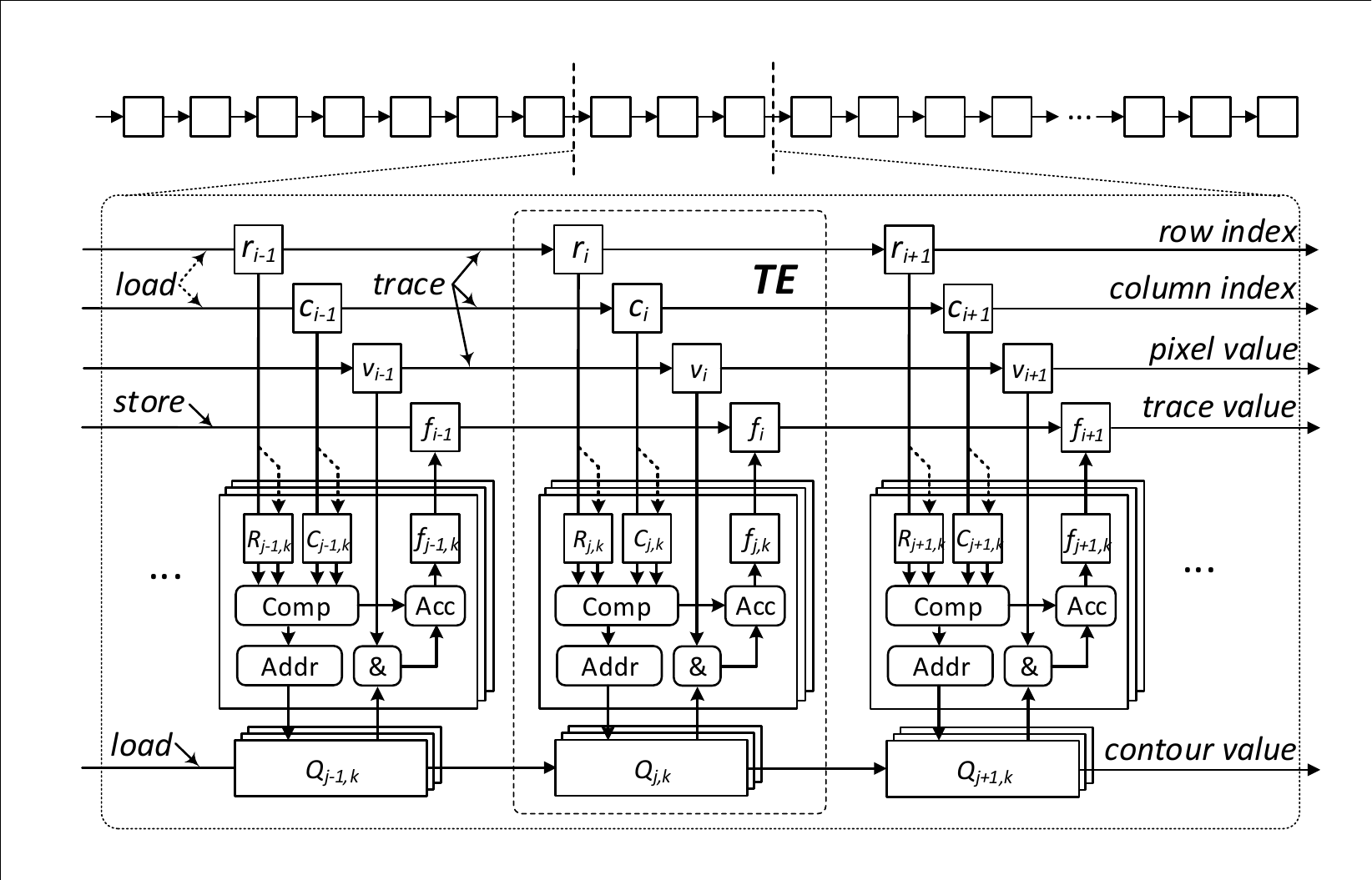}}
\caption{ACC-Trace: Efficient calcium trace extraction with 1-D systolic array architecture \cite{Chen2021}.}
\label{fig:4}
\end{figure}

\begin{equation} \label{eq:1}
  {{f}_{j,k}}=\sum\limits_{i=0}^{N}{\sum\limits_{d{{r}_{ijk}}=0}^{{{N}_{C}}}{\sum\limits_{d{{c}_{ijk}}=0}^{{{N}_{C}}}{{{v}_{i}}\cdot {{Q}_{j,k}}\left( d{{r}_{ijk}},d{{c}_{ijk}} \right)}}}.
\end{equation}

The (${dr}_{ijk}$,${dc}_{ijk}$) are derived indices from the scanned indices (${r}_{i}$,${c}_{i}$) and the cell center location (${R}_{j,k}$,${C}_{j,k}$):

\begin{equation} \label{eq:2}
\left\{ \begin{matrix}
   d{{r}_{ijk}}={{r}_{i}}-{{R}_{j,k}}+{{N}_{C}}/2  \\
   d{{c}_{ijk}}={{c}_{i}}-{{C}_{j,k}}+{{N}_{C}}/2  \\
\end{matrix} \right..
\end{equation}

The ACC-Trace accelerator operates under 3 modes. 1) Load: Cell contour and center location information are loaded to the local cell contour buffer and local register files in a bit-shift fashion. 2) Compute: Calcium image is scanned pixel by pixel through the chain of TEs, whereas the TEs work in parallel to monitor the scanned pixels and accumulate trace values based on the contour information fetched correspondingly for each pair of incoming row and column indices (${r}_{i}$,${c}_{i}$). 3) Store: The accumulated trace values ${f}_{jk}$ are sent to an external buffer in the same bit-shift fashion. Though the ACC-Trace design is very scalable, oftentimes we are not able to build a sufficiently long chain of TEs to handle the trace extraction tasks for all target cells due to limited FPGA resources at hand. In this case, we build a relatively shorter ACC-Trace chain and reuse it for multiple iterations of operation, and reload the cell contour and center information between iterations.

\subsection{Cell-Based Contour Allocation} \label{sec:3.2}

In the ACC-Trace accelerator design, each TE stores $K$ contours in its local buffer. We implement the local buffer by a single BRAM slice on the FPGA, and use only one read port for contour access during the trace extraction given the limited number of ports provided by the BRAM. This will raise a memory access conflict when the TE tries to fetch data from two different contours at the same clock cycle. In order to avoid the memory access conflict, we need to guarantee that there is no overlap among $K$ stored contours in every TE on the algorithm level.

\begin{figure}
\centerline{\includegraphics[width=3.3in]{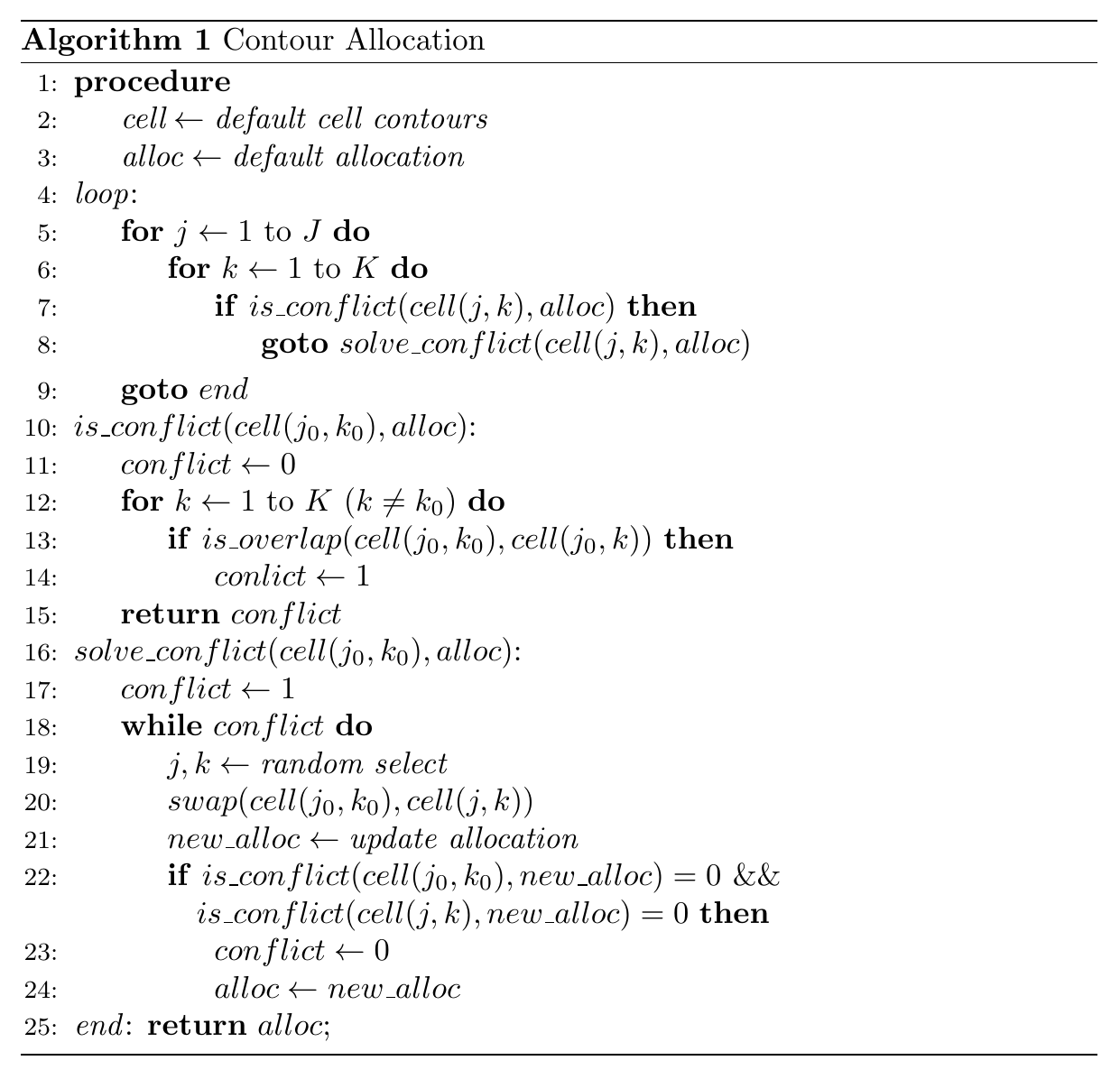}}
\caption{Algorithm for allocating contours to the ACC-Trace accelerator.}
\label{fig:5}
\end{figure}

We propose a cell contour allocation algorithm to make sure the cells assigned to a same TE do not overlap, as shown in Fig. \ref{fig:5}. Initially, we allocate ${J}\times{K}$ cell contours to $J$ TEs with a default order. Then we loop through each contour and check if the current contour of Cell($j$,$k$) in a TE has overlap with other contours allocated in the same TE. If it has an overlap with another contour, then it means that there will be a memory access conflict at this contour, and the algorithm will go to a \verb|solve_conflict| routine to adjust the allocation for this contour. In the \verb|solve_conflict| routine, the algorithm will iteratively pick up a random contour from another TE, swap the conflict contour of Cell(${j}_{0}$,${k}_{0}$) and the current contour of Cell($j$,$k$), and check if there is any overlap between these two contours and the reset of contours in the same TE after swapping. If there is no overlap, the algorithm will accept the swap as a solution for the conflict and update the contour allocation correspondingly. Otherwise, the algorithm will go on to pick up another random contour and go through the same checking described by line 20-24 in Algorithm 1 shown in Fig. 5 until it finds an acceptable solution.

\subsection{Tile-Based Contour Allocation} \label{sec:3.3}

\begin{figure}
\centerline{\includegraphics[width=3.5in]{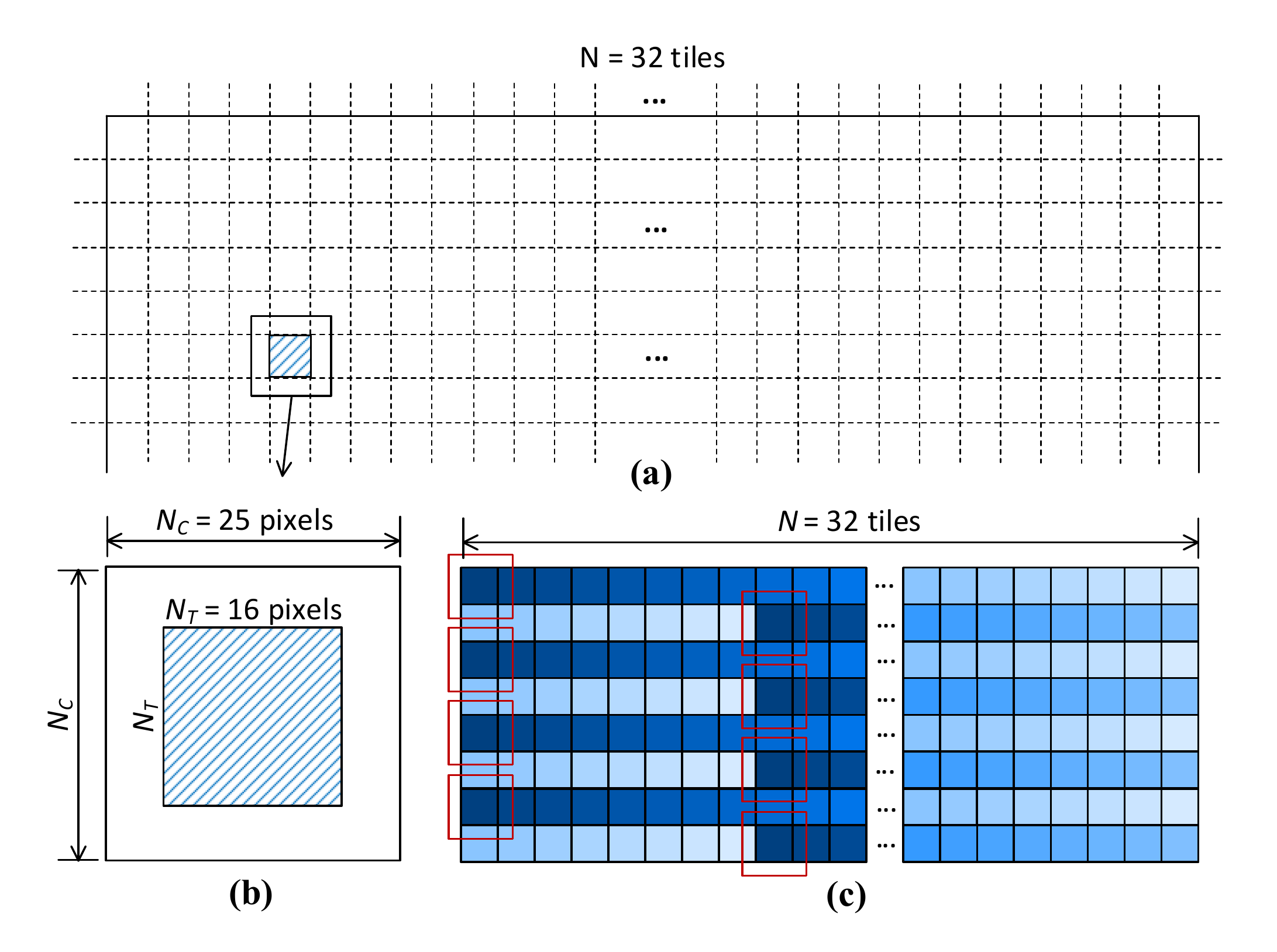}}
\caption{(a) Tile based contour and (b) allocation of tile-based contours to the ACC-Trace accelerator.}
\label{fig:6}
\end{figure}

Fig. \ref{fig:6}(a) and (b) illustrate the concept of the tile-based contours. As the image size is $512\times 512$, we can divide them evenly into ${N}\times{N}$ ($N$ = 32) non-overlapping ${{N}_{T}}\times{{N}_{T}}$ (${{N}_{T}}$ = 16) tiles as shown in Fig. \ref{fig:6}(a), and each tile can be viewed as a ${{N}_{C}}\times{{N}_{C}}$ (${{N}_{C}}$ = 25) tile-based contour. Inside each tile-based contour, we set the elements at the central $16\times 16$ region to be ``1'', while keeping the rest of elements to be ``0''. By taking advantage of the tile-based contours for the calcium trace extraction, we can eliminate the need of running offline analysis pipeline, such as CaImAn \cite{Giovannucci2019}, for cell contour detection. The offline analysis pipeline usually takes tens of minutes or even longer to finish on a pre-recorded session of calcium images depending on the computation platform the pipeline runs on. With tile-based contours, we can bypass the cell contour detection step and directly extract calcium traces, and it contributes to a short turnaround time for decoding model training and a quick deployment of decoder onto the hardware for closed-loop feedback applications \cite{Chen2022_eLife}. In Section \ref{sec:4}, we will provide more evaluation results regarding the comparison between the decoding based on the tile-based and cell-based contours.

Based on the ACC-Trace accelerator microarchitecture introduced in Section 3, a direct mapping of the $32\times 32$ tile-based contours to the $J$ = 32 accelerator will cause memory access conflict, as the neighboring tile-based contours mapped to the same TE have overlaps with each other.  As Fig. \ref{fig:6}(c) shows, we come up with a way of mapping of tile-based contours in order to avoid memory access conflicts. Here, 8 tile-based contours highlighted by $25\times 25$ size squares are shown with the same shaded color and mapped to the same TE, and following tile-based contours in these rows are mapped to following TEs, so on and so forth. In this way, the tile-based contours mapped to every single TE do not have any overlap, so the memory access conflicts are avoided during the scanning of pixels in the trace extraction computation.

\subsection{Latency Optimizations} \label{sec:3.4}

\begin{figure*}
\includegraphics[width=\textwidth]{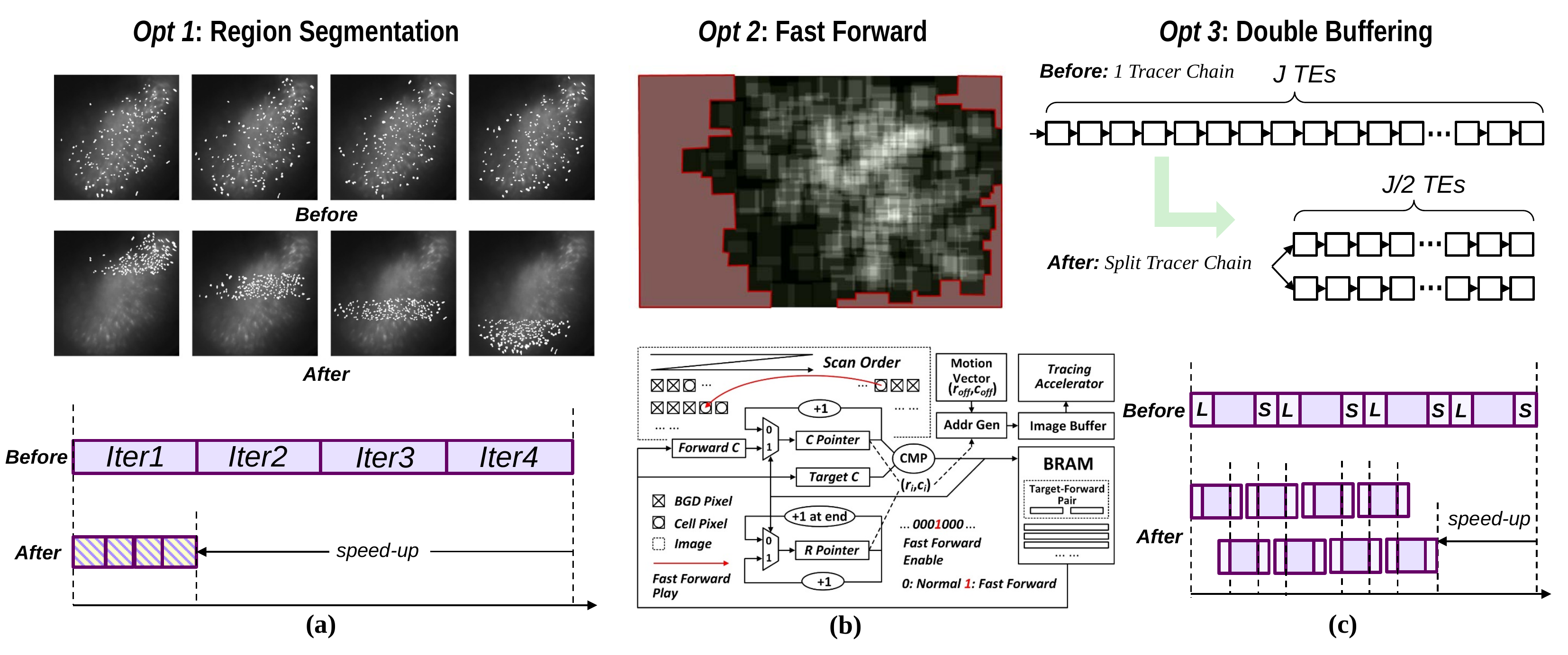}
\caption{Summary of proposed latency optimizations for the ACC-Trace accelerator: (a) Region segmentation; (b) Fast forward; (c) Double buffering.}
\label{fig:7}
\end{figure*}

As Fig. \ref{fig:3} shows, the runtime of the trace extraction step largely contributes to the overall processing and decoding latency. In this subsection, we introduce three optimizations that can speed up the trace extraction and reduce the overall latency, as shown in Fig. \ref{fig:7}.

Fig. \ref{fig:7}(a) presents the first latency optimization: region segmentation. This optimization takes effect when the ACC-Trace accelerator needs to be reused for multiple times during the trace extraction of a single frame. If we do not consider the region segmentation, the contours that need to be processed at each round are distributed evenly across the image. In that case, the whole frame of the image needs to be scanned in order to complete the trace extraction task. On the contrary, if we constrain the contours by their locations when allocating them to different rounds of trace extraction, we only need to scan a part of the image at each round. In this way, we can largely save runtime without increasing the overhead on the hardware implementation.

Within one iteration, we find that there is more opportunity for the latency optimization, as Fig. \ref{fig:7}(b) shows. While a portion of pixels in the image are in the background and do not have a cover of the cell footprint, scanning these pixels simply adds up runtime and but does not make an effect on the trace extraction results. The main idea of our second latency optimization is to skip over those pixels in the scanning. Through an offline analysis of the locations of the cell contours, we can figure out starting and ending position indices of each consecutive segment of background pixels. We store these position indices in a local BRAM buffer aside the ACC-Trace accelerator and design digital circuits that keep monitoring the indices of scanned pixels and take action of the fast forward when a background pixel is detected. 

The third latency optimization is the double buffering as shown in Fig. \ref{fig:7}(c). Suppose we design an ACC-Trace accelerator with $J$ TEs. By default, the load, compute and store operate in sequential order as the same data and indices transmission paths are shared among operations. If we consider splitting the accelerator chain into two parts, we can double buffer the operation of these two parts, and hide the load and store time behind the compute operation without incurring overhead on the hardware implementation. In this way, additional speed-up can be achieved for further reduced latency.

Taking a combination of the proposed latency optimizations, the ACC-Trace accelerator can reduce the runtime from 3.5 ms to 589 $\mu$s on an evaluation set with 760 cell contours under a 300 MHz clock frequency, whereas the increased LUT, FF and BRAM resource usage is within 1.5\%.

\section{ACC-Decode: Efficient Decoding} \label{sec:4}

In this section, we first review the position decoding task \cite{Chen2022_eLife}, and then discuss the proposed neural network based decoding methods and accelerator designs, which take input from the extracted traces by the ACC-Trace accelerator. 

\subsection{Position Decoding Task} \label{sec:4.1}

The task is to decode rat's positions on a 2.5-m linear track as it runs back and forth \cite{Chen2022_eLife, Chen2022_ISCAS}. The linear track is evenly divided into 24 bins, labeled with discrete numbers 0, 1, 2, ..., 23. Since the real-time decoding aims at closed-loop feedback applications, it is required to be causal, which means that the decoding does not rely on future calcium images as input.

\subsection{Accuracy Evaluation Metrics} \label{sec:4.2}

We came up with two accuracy evaluation metrics for the position decoding task: 1) \textit{Hit-N} accuracy measurement, and 2) mean error measurement, as elaborated below.

\subsubsection{\textit{Hit-N} Accuracy} \label{sec:4.2.1}

The \textit{Hit-N} accuracy counts the percentage of frames for which the decoded position falls into the \textit{N}-sample neighborhood around the true position. For example, the \textit{Hit-1} accuracy measures the percentage of frames for which the decoded position is exactly right, whereas the \textit{Hit-3} accuracy measures the percentage of frames for which the decoded position falls into the $\pm$1-bin neighborhood around the true position.

\subsubsection{Mean Error} \label{sec:4.2.2}

The mean error $\sigma$ measures the absolute difference between the decoded position and the true position on average for each frame. It reflects the decoding accuracy in general and can be used as a fair comparison metric across different decoding methods on the same dataset.

\subsection{Decoding Methods} \label{sec:4.3}

\begin{figure}
\centerline{\includegraphics[width=3.5in]{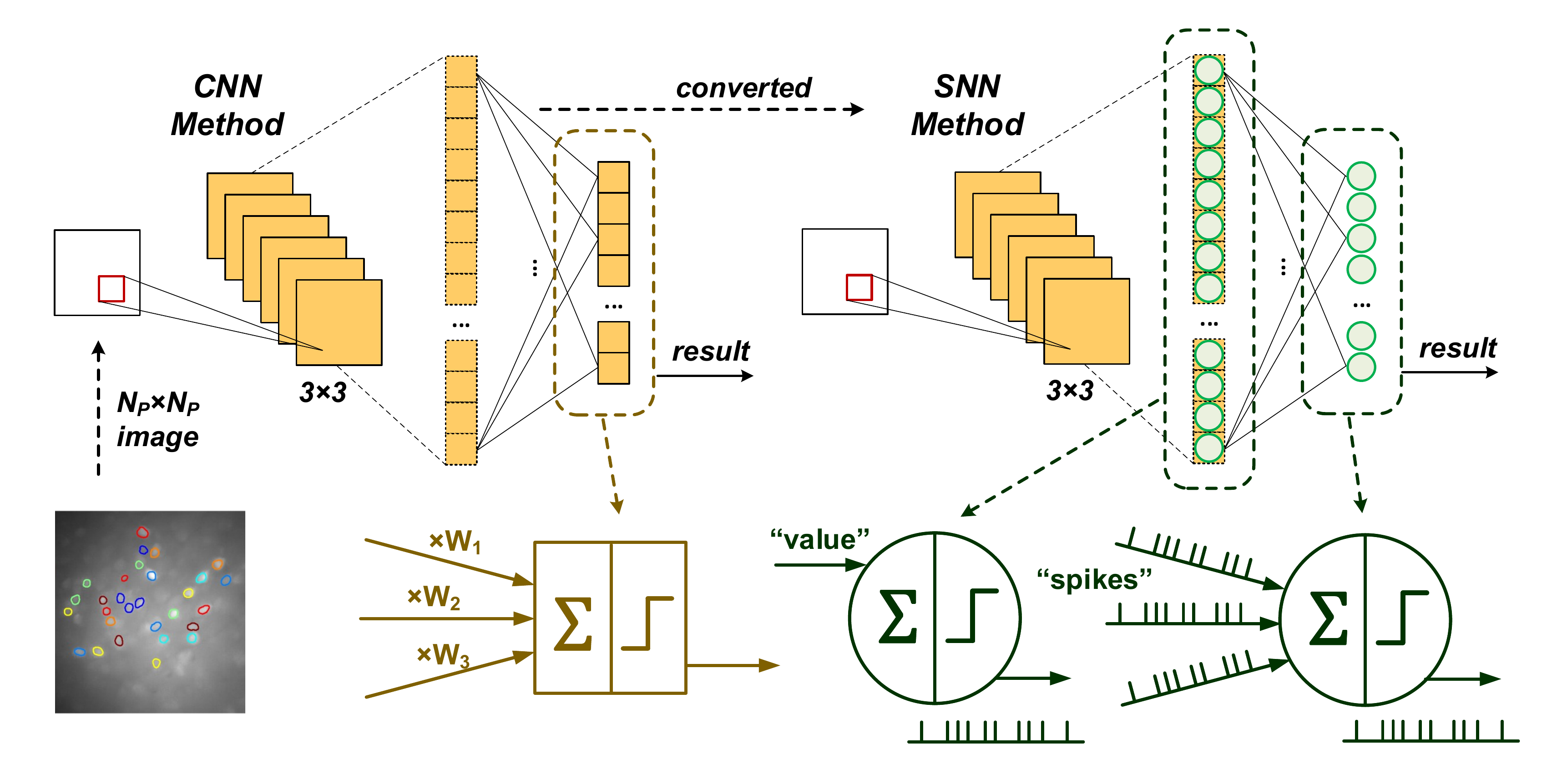}}
\caption{Proposed CNN and SNN based methods for position decoding from calcium images.}
\label{fig:8}
\end{figure}

We first propose the CNN and the SNN based decoding methods for the position decoding. Fig. \ref{fig:8} shows the proposed CNN and SNN models, and they share the same model architecture that we decide to use based on the algorithm exploration on the decoding task \cite{Chen2022_ISCAS}. For the calcium image decoding, we first extract all of the cells from the calcium images in the training set offline using the CaImAn method \cite{Giovannucci2019}. Then we identify the first set of ${{N}_{P}}\times{{N}_{P}}$ cells with a largest possible ${N}_{P}$ from the extracted cells based on a downwards peak trace value sorting [15], normalize the trace values collected from these cells to [0,1], and form an image as the CNN and SNN inputs. For the CNN model, the input image is first convolved with 6 filter kernels in 3$\times$3 size, then the feature maps spread out to an $({N}_{P}-2)\times({N}_{P}-2)\times6$ element vector. A hidden layer establishes all-to-all connections from these elements to 24 output nodes, and each output node corresponds to a specific position bin on the linear track. Finally, the accumulated values at all output nodes are compared, and the node that has the maximum value is identified and the corresponding position label is generated as the decoding result.

Our proposed SNN decoder adopts a rate-based encoding, and it is converted from the CNN by the SNN Toolbox \cite{Rueckauer2017}. The conversion relies on an integrate-and-fire (IAF) neuron model. As Fig. \ref{fig:8} illustrates, the input value is accumulated to a potential at each hidden node at every time step. Whenever the potential passes a threshold ${V}_{t}$, the hidden node (as a neuron) produces an output of ``1'' (as a spike) at a specific time step, and a value of ${V}_{t}$ is subtracted from the potential. The generated spikes at the hidden layer propagate to the output layer. The output nodes are also modeled as IAF neurons. Each output node takes input spikes from $({N}_{P}-2)\times({N}_{P}-2)\times6$ IAF neurons in the hidden layer, and produces output spikes under the same threshold ${V}_{t}$. Whenever an output node receives an input spike from a specific cell, a constant weight bound with the connection is accumulated in the same fashion as the hidden node. Finally, the number of spikes generated by each output node is counted over the number of time steps TS for each SNN inference. The output node that produces the maximum number of spikes is identified and its index serves as the decoding outcome.

We then propose a simple two-hidden-layer ANN model for the same position decoding task. Each hidden layer contains 32 nodes. We explored two different encodings for the output layer: 1) Categorical encoding, which contains 24 nodes with each node representing a specific position on the linear track; 2) Ordinal encoding, which contains 12 nodes. The inference algorithm converts the 12 output node values into a 12-bit binary code by setting a threshold of 0.5 at each output node. The number of consecutive 1s or 0s from the left of the 12-bit binary code indicates the decoded position bin index. We evaluated effects of 1) the decoding option: tile-based/cell-based decoding and 2) the encoding method: categorical/ordinal encoding on the decoding accuracy across datasets collected from 6 different rats.

\subsection{ACC-Decode: Efficient Decoder} \label{sec:4.4}

\begin{figure}
\centerline{\includegraphics[width=3.3in]{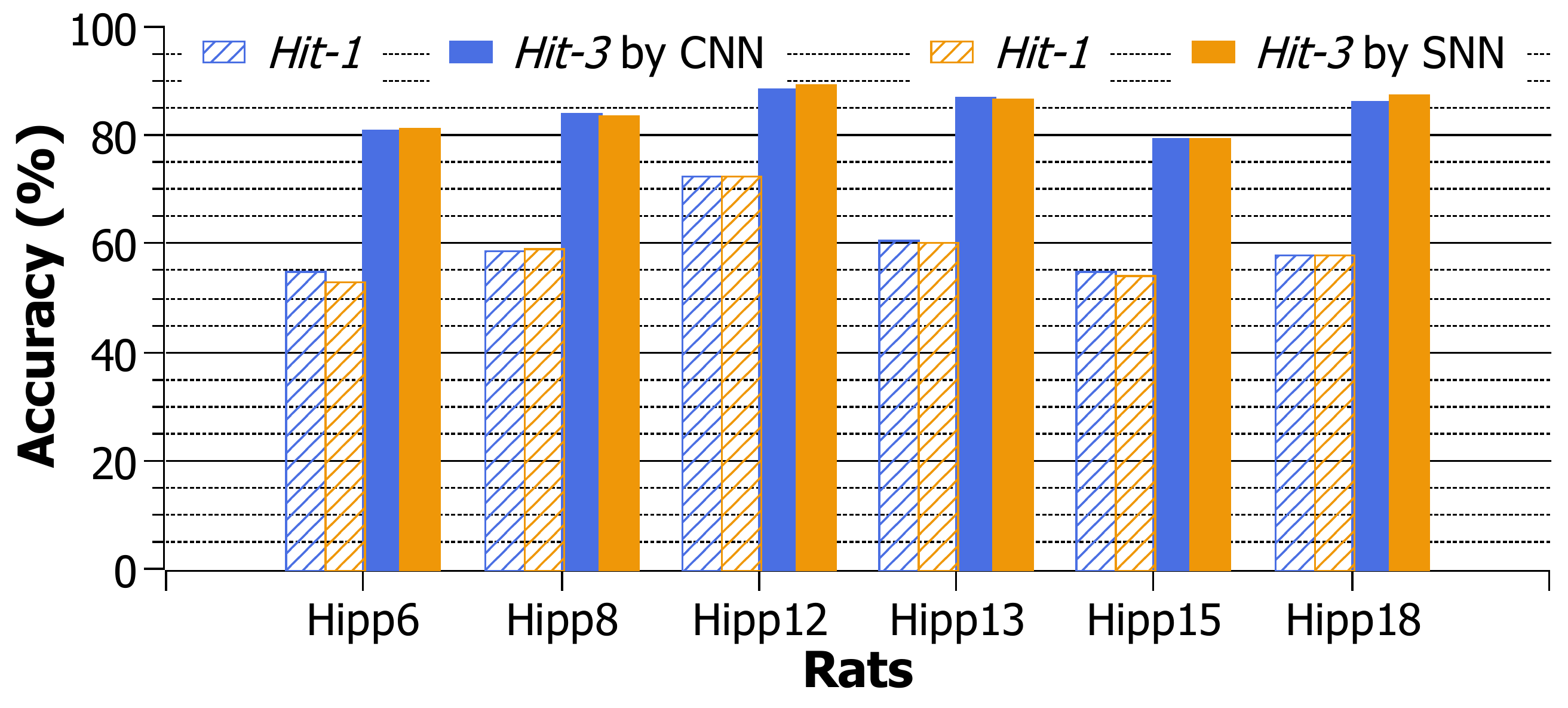}}
\caption{\textcolor{blue}{\textit{Hit-1}/\textit{Hit-3} accuracy evaluated for CNN and converted SNN decoders on 6 different rats.}}
\label{fig:9}
\end{figure}

We evaluated the \textit{Hit-1}/\textit{Hit-3} accuracy achieved by a baseline floating-point CNN and a converted 8-bit SNN with ${T}_{S}$ = 8 time steps on calcium image decoding test sets across 6 different rats. Table 1 presents an overview of these test sets. Among 6 rats ${R}_{1}$-${R}_{8}$, there were 153-760 cells detected from their corresponding training sets. According to Section \ref{sec:4.2}, we assigned ${N}_{P}$ with the value of 13-27 for these rats. We collected 8000 frames of calcium images with corresponding tracking positions for each rat, and divided each recording session into two parts for training and testing separately. \textit{Hit-1}/\textit{Hit-3} accuracies on the test sets for the CNN and the SNN models are evaluated and summarized in Fig. \ref{fig:9}. On average, the CNN and SNN models achieve 56.3\%/83.1\% and 56.0\%/82.8\% \textit{Hit-1}/\textit{Hit-3} accuracy, correspondingly. It can be observed that the 8-bit SNN with 32 time steps has almost no accuracy loss compared to the floating-point CNN in performing the decoding task across rats.

\begin{table}
  \caption{The Numbers of Cells Identified and Selected for Decoding}
  \label{tab:1}
  \centering
  \begin{tabular}{|l|c|c|c|c|c|c|}
    \hline
    Rat's ID & Hipp6 & Hipp8 & Hipp12 & Hipp13 & Hipp15 & Hipp18\\
    \hline
    \# Cells & 296 & 309 & 317 & 194 & 760 & 643\\
    \hline
    ${N}_{P}\times{N}_{P}$ & 16$\times$16 & 17$\times$17 & 17$\times$17 & 13$\times$13 & 27$\times$27 & 25$\times$25\\
    \hline
  \end{tabular}
\end{table}

\begin{figure}
\centerline{\includegraphics[width=3.6in]{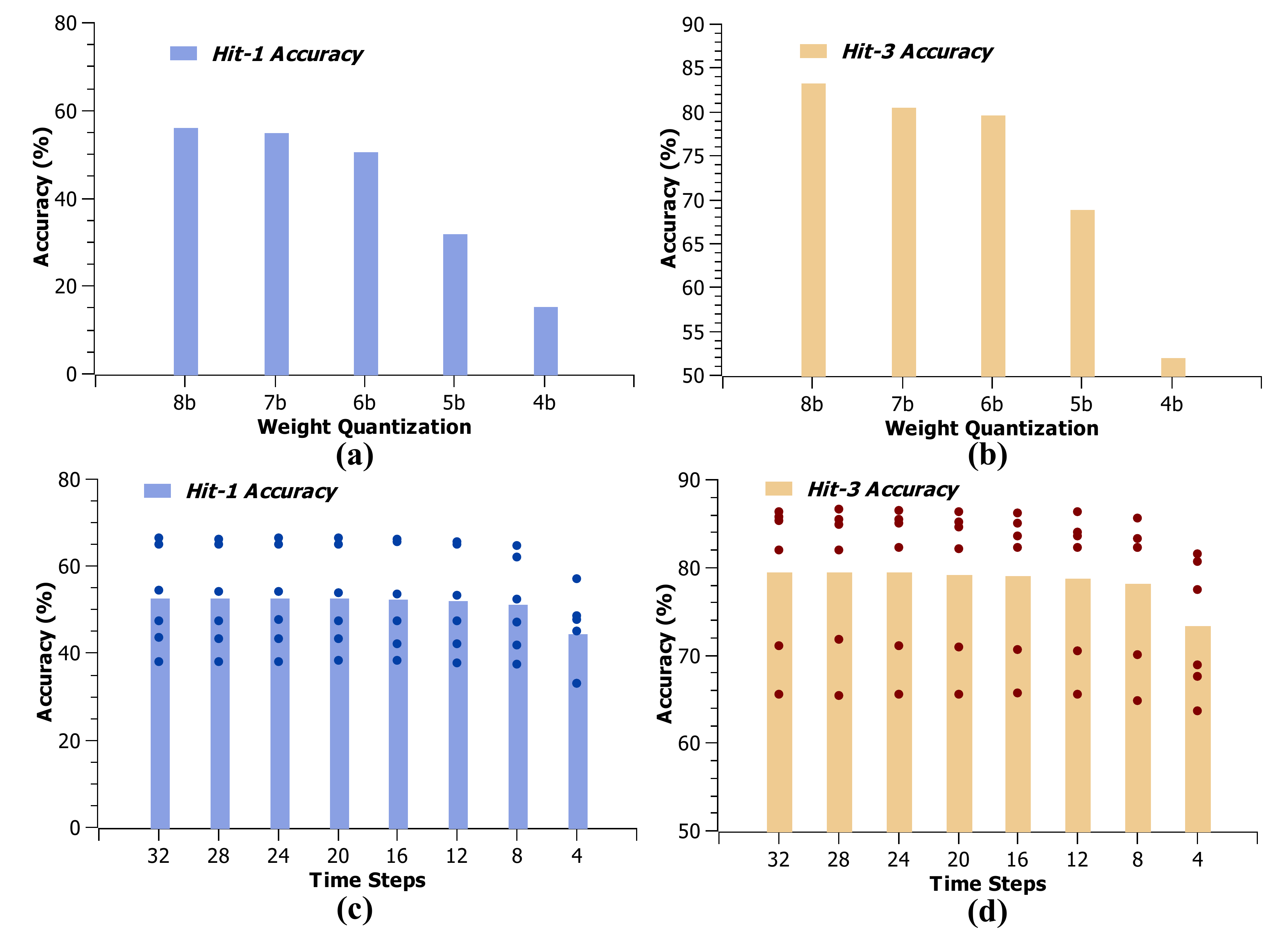}}
\caption{Accuracy evaluation for the SNN based method under various quantization of weights and inference time steps: (a) and (b) show the \textit{Hit-1}/\textit{Hit-3} accuracy for a 32 time-step SNN under different bit-widths of weights. (c) and (d) show the \textit{Hit-1}/\textit{Hit-3} accuracy for a 6-bit SNN with different inference time steps.}
\label{fig:10}
\end{figure}

We further apply fixed-point quantization and time-step reduction for the SNN model. Fig. \ref{fig:10} shows the accuracy evaluation results under various weight quantization and time steps. We first keep the time step constant and gradually apply more aggressive bit-width quantization. From Fig. \ref{fig:10}(a), we see that the SNN model under 6-bit quantization on average has 5.58\%/3.62\% loss on the \textit{Hit-1}/\textit{Hit-3} accuracy. Then we apply the time-step reduction for the 6-bit SNN model, and according to Fig. \ref{fig:10}(b), we can reduce time steps of the SNN from 32 to 16, while incurring less than 1\% \textit{Hit-1} and \textit{Hit-3} accuracy loss. Table \ref{tab:2} shows the comparison on FPGA resource usage among different decoder models. We set the input size to be the same at $16\times 16$ for all these models. The 6-bit SNN consumes less LUT, FF, and BRAM, and completely avoids the usage of DSP compared to the 8-bit CNN model. The ANN model consumes higher LUT and FF usage, but saves the BRAM resource against the CNN model. Under 300 MHz clock frequency, the cycle counts for the CNN, SNN (considering 8 time steps), and ANN models are 65.9k, 278.9k, and 19.9k, respectively.

\begin{table}
  \caption{FPGA Resource Usage by Decoding Accelerator Kernels}
  \label{tab:2}
  \centering
  \begin{tabular}{|l|c|c|c|}
    \hline
    & CNN-Based & SNN-Based & ANN-Based\\
    \hline
    LUT & 2713 & 1484 & 3356\\
    \hline
    FF & 2454 & 1440 & 3133\\
    \hline
    DSP & 7 & 0 & 7\\
    \hline
    BRAM & 18 & 14 & 6\\
    \hline
  \end{tabular}
\end{table}

Table \ref{tab:3} shows a more comprehensive decoding performance evaluation by the Hit-3 accuracy and the mean error $\sigma$ on the decoded position for the proposed CNN/ANN based decoding methods under various input (cell-based vs. tile-based) and output (categorical vs ordinal) settings. The evaluation results are derived from the average of 5 independent experimental trials. We make several observations from the evaluation results: 1) the tile-based decoding generally outperforms the cell-based decoding, which is consistent with the finding in \cite{Chen2022_eLife}. This can be explained by the fact that more tile-based contours (900 vs. 169-729) are involved in the decoding. 2) The ANN based method slightly outperforms the CNN counterpart. That's because the extracted trace data does not contain visual features that the CNN can leverage. On the contrary, the ANN can take the advantage of its global connections for more efficient and accurate decoding. 3) The ordinal encoding in general outperforms the categorical encoding. That's because the decoded positions are continuous, and the ordinal encoding inherently suits this characteristic.

We designed CNN/ANN based decoding accelerator kernels with the Vitis HLS for the cell-based scenario. Table \ref{tab:4} shows hardware resource utilization from the post implementation report. Compared to the CNN based design, the ANN based design consumes 30-35\% more LUT and FF, and much less BRAM resources. Table \ref{tab:5} reports the runtime evaluated on the CNN/ANN based decoding kernels. As we targeted 300 MHz for the decoding kernels, the runtime for the CNN-based decoding kernel ranges from 155-800 $\mu$s, while the ANN-based decoding kernel takes much shorter runtime by a factor of 6.5-9.5x.

\begin{table*}
  \caption{Accuracy Evaluation for CNN/ANN Based Decoding Methods}
  \label{tab:3}
  \centering
  \begin{tabular}{|l|c|c|c|c|c|c|c|c|c|c|c|c|}
    \hline
    Metric & \multicolumn{6}{|c|}{\textit{Hit-3} Accuracy (\%)} & \multicolumn{6}{|c|}{Mean Error $\sigma$}\\
    \hline
    Input & \multicolumn{3}{|c|}{Cell-Based} & \multicolumn{3}{|c|}{Tile-Based} & \multicolumn{3}{|c|}{Cell-Based} & \multicolumn{3}{|c|}{Tile-Based}\\
    \hline
    Output & \multicolumn{2}{|c|}{Categorical} & Ordinal & \multicolumn{2}{|c|}{Categorical} & Ordinal & \multicolumn{2}{|c|}{Categorical} & Ordinal & \multicolumn{2}{|c|}{Categorical} & Ordinal\\
    \hline
    Model & CNN & \multicolumn{2}{|c|}{ANN} & CNN & \multicolumn{2}{|c|}{ANN} & CNN & \multicolumn{2}{|c|}{ANN} & CNN & \multicolumn{2}{|c|}{ANN}\\
    \hline
    Hipp6 & 82.69 & 87.93 & 89.88 & 86.72 & 90.66 & \textbf{92.39} & 1.800 & 0.736 & 0.618 & 0.888 & 0.563 & \textbf{0.508}\\
    \hline
    Hipp8 & 82.87 & 86.89 & 88.99 & 88.49 & 90.40 & \textbf{91.39} & 1.091 & 0.723 & 0.621 & 0.597 & 0.579 & \textbf{0.553}\\
    \hline
    Hipp12 & 88.97 & 95.37 & 95.29 & 94.07 & 96.73 & \textbf{97.30} & 0.667 & 0.291 & 0.285 & 0.386 & 0.236 & \textbf{0.226}\\
    \hline
    Hipp13 & 86.28 & 96.30 & 94.97 & 92.07 & \textbf{98.55} & 96.64 & 0.877 & 0.338 & 0.423 & 0.460 & \textbf{0.252} & 0.593\\
    \hline
    Hipp15 & 78.69 & 83.47 & 82.84 & 78.10 & \textbf{87.43} & 87.41 & 1.710 & 0.924 & 0.890 & 2.220 & \textbf{0.782} & 0.846\\
    \hline
    Hipp18 & 86.67 & 80.89 & 83.71 & \textbf{86.83} & 79.68 & 86.26 & 1.181 & 1.167 & \textbf{0.779} & 1.081 & 1.352 & 0.911\\
    \hline
  \end{tabular}
\end{table*}

\begin{table*}
  \caption{Hardware Resource Usage for Cell-Based CNN/ANN Based Decoding Accelerator Kernels}
  \label{tab:4}
  \centering
  \begin{tabular}{|l|c|c|c|c|c|c|c|c|c|c|c|c|}
    \hline
    Resource & \multicolumn{3}{|c|}{LUT} & \multicolumn{3}{|c|}{FF} & \multicolumn{3}{|c|}{DSP} & \multicolumn{3}{|c|}{BRAM}\\
    \hline
    Output & \multicolumn{2}{|c|}{Categorical} & Ordinal & \multicolumn{2}{|c|}{Categorical} & Ordinal & \multicolumn{2}{|c|}{Categorical} & Ordinal & \multicolumn{2}{|c|}{Categorical} & Ordinal\\
    \hline
    Model & CNN & \multicolumn{2}{|c|}{ANN} & CNN & \multicolumn{2}{|c|}{ANN} & CNN & \multicolumn{2}{|c|}{ANN} & CNN & \multicolumn{2}{|c|}{ANN}\\
    \hline
    Hipp6 & 2,713 & 3,394 & 3,472 & 2,454 & 3,236 & 3,297 & 7 & 7 & 7 & 18 & 6 & 6\\
    \hline
    Hipp8/12 & 2,525 & 3,411 & 3,529 & 2,460 & 3,250 & 3,311 & 7 & 7 & 7 & 19 & 8 & 8\\
    \hline
    Hipp13 & 2,589 & 3,390 & 3,475 & 2,427 & 3,239 & 3,298 & 16 & 11 & 11 & 12 & 6 & 6\\
    \hline
    Hipp15 & 2,526 & 3,445 & 3,523 & 2,492 & 3,270 & 3,333 & 16 & 11 & 11 & 54 & 14 & 14\\
    \hline
    Hipp18 & 2,536 & 3,415 & 3,522 & 2,492 & 3,276 & 3,341 & 16 & 11 & 11 & 54 & 12 & 12\\
    \hline
  \end{tabular}
\end{table*}

\begin{table}
  \caption{Runtime for CNN/ANN Based Decoding Accelerator Kernels}
  \label{tab:5}
  \centering
  \begin{tabular}{|l|c|c|c|c|c|c|}
    \hline
    Metric & \multicolumn{3}{|c|}{Cycle Count} & \multicolumn{3}{|c|}{Runtime ($\mu$s)}\\
    \hline
    Output & \multicolumn{2}{|c|}{Categorical} & Ordinal & \multicolumn{2}{|c|}{Categorical} & Ordinal\\
    \hline
    Model & CNN & \multicolumn{2}{|c|}{ANN} & CNN & \multicolumn{2}{|c|}{ANN}\\
    \hline
    Hipp6 & 65,936 & 10,196 & 9,840 & 219.8 & 34.0 & 32.8\\
    \hline
    Hipp8/12 & 77,030 & 11,253 & 10,897 & 256.8 & 37.5 & 36.3\\
    \hline
    Hipp13 & 46,553 & 7,417 & 7,061 & 155.2 & 24.7 & 23.5\\
    \hline
    Hipp15 & 240,089 & 25,337 & 24,981 & 800.3 & 84.5 & 83.3\\
    \hline
    Hipp18 & 203,225 & 22,009 & 21,653 & 677.4 & 73.4 & 72.2\\
    \hline
  \end{tabular}
\end{table}

\section{Implementation} \label{sec:5}

\begin{figure}
\centerline{\includegraphics[width=3.0in]{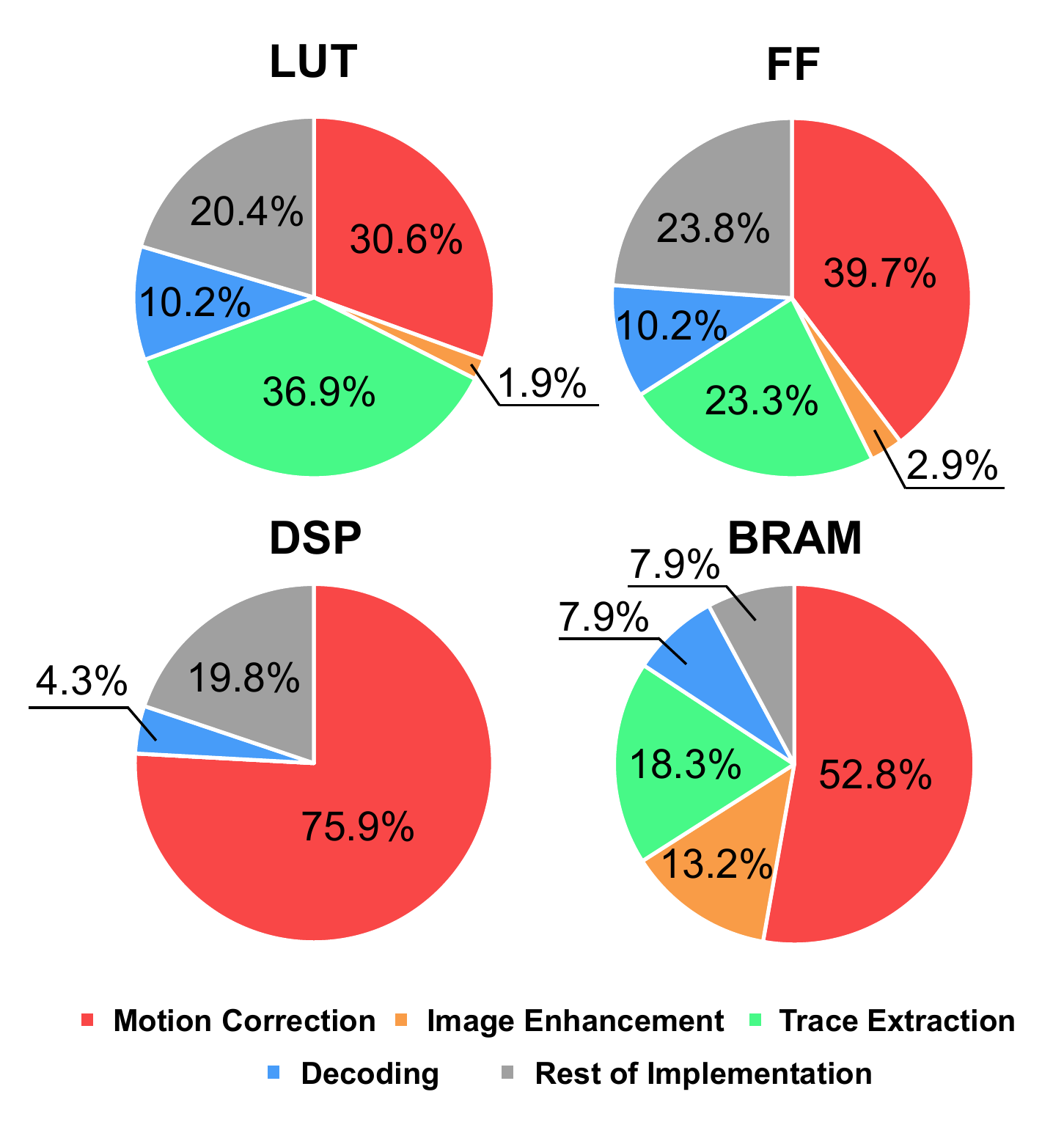}}
\caption{Breakdown of the FPGA resource usage of the implementation for the real-time calcium image processing and decoding.}
\label{fig:11}
\end{figure}

We implemented the proposed calcium image processing pipeline introduced in Session \ref{sec:2} on the Ultra96 SoC platform. The complete processing pipeline operates at 300 MHz. Table \ref{tab:6} reports the overall FPGA resource utilization, and Fig. \ref{fig:11} shows the resource usage breakdown for the implemented accelerators. The motion correction consumes the largest part of computation and memory resources, as it speeds up the most time-consuming and critical pre-processing step, which removes motion artifacts from the raw calcium images and helps increase the \textit{Hit-3} decoding accuracy by 1.42\% on average based on our analysis with the CNN-based decoder. The ACC-Trace accelerator also costs considerable LUT, FF and BRAM resources given its fine-grain pipeline architecture. It largely reduces the input dimension and inference time of the decoder whereas it does not increase accuracy loss. The ACC-Decode accelerator consumes less resources than the ACC-Trace accelerator, and the remaining resource is reserved for implementing the feedback control logic on the same FPGA. 

We measured the runtime of the proposed neural network based decoders on the embedded ARM processor of the Ultra96 platform under 1 GHz operating frequency. The runtime of the CNN and ANN based decoders with the 27×27 input size are 7.45 ms and 1.42 ms per inference, respectively. The runtime of the SNN based decoder with the 16×16 input size and 8 time steps is 7.65ms per inference. Compared to the ARM based implementation, the FPGA implementation of the CNN, ANN, and SNN based decoders achieve 9.6x, 17.1x, and 8.2x speedup, respectively.

\begin{table}
  \caption{The Overall FPGA Resource Utilization.}
  \label{tab:6}
  \centering
  \begin{tabular}{|l|c|c|c|}
    \hline
    Resource & Utilization & Availability & Utilization \%\\
    \hline
    LUT & 52301 & 70560 & 74.12\%\\
    \hline
    FF & 60490 & 141120 & 42.86\%\\
    \hline
    DSP & 116 & 360 & 32.22\%\\
    \hline
    BRAM & 216 & 216 & 100\%\\
    \hline
  \end{tabular}
\end{table}

We built a customized hardware interface board that can be plugged on top of the Ultra96 platform, as Fig. 12 shows. The interface board provides the input port for connecting the miniscope sensor, the output port for applying feedback stimulation with TTL pulses, and a standard Ethernet port that can communicate with the host computer for data transfer and user interaction. The input port is connected to the DAQ board with 13 bundled fly wires, which correspond to one 66.67 MHz sensor readout clock, 8 sensor data pins, H/V synchronization signals, and VDD/GND signals. We also connect the DAQ board to the host computer with a USB 3.0 cable to control the miniscope sensor, and connect the miniscope sensor to the DAQ board with a 1.5-m flexible coax cable. The overall system except the host computer is powered by a 12V power supply. The peak power consumption is 5.3W, and the standby power consumption is 2.2W.

\begin{figure}
\centerline{\includegraphics[width=2.5in]{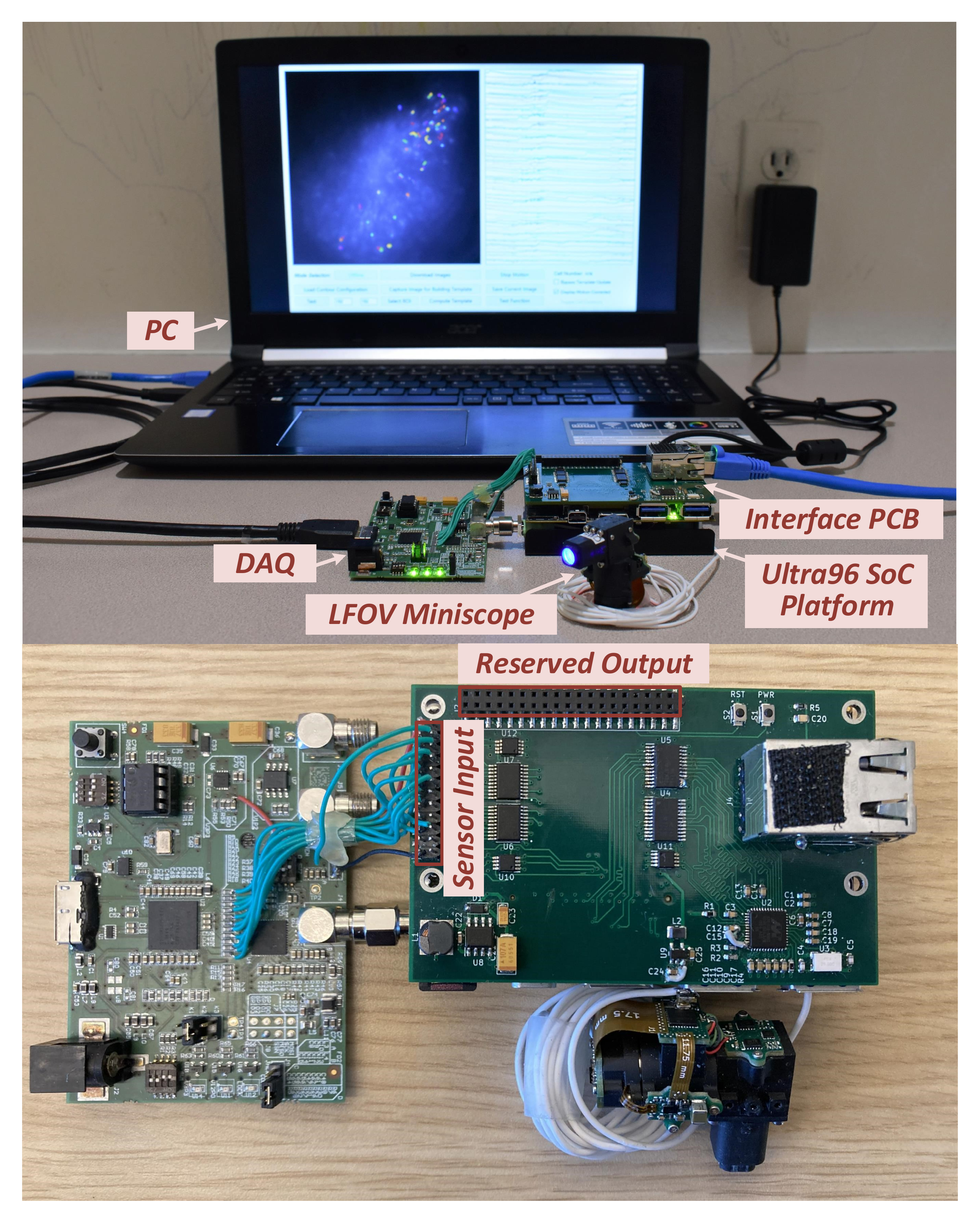}}
\caption{The hardware interface and the demonstration setup for the closed-loop feedback system.}
\label{fig:12}
\end{figure}

\begin{figure*}
\includegraphics[width=\textwidth]{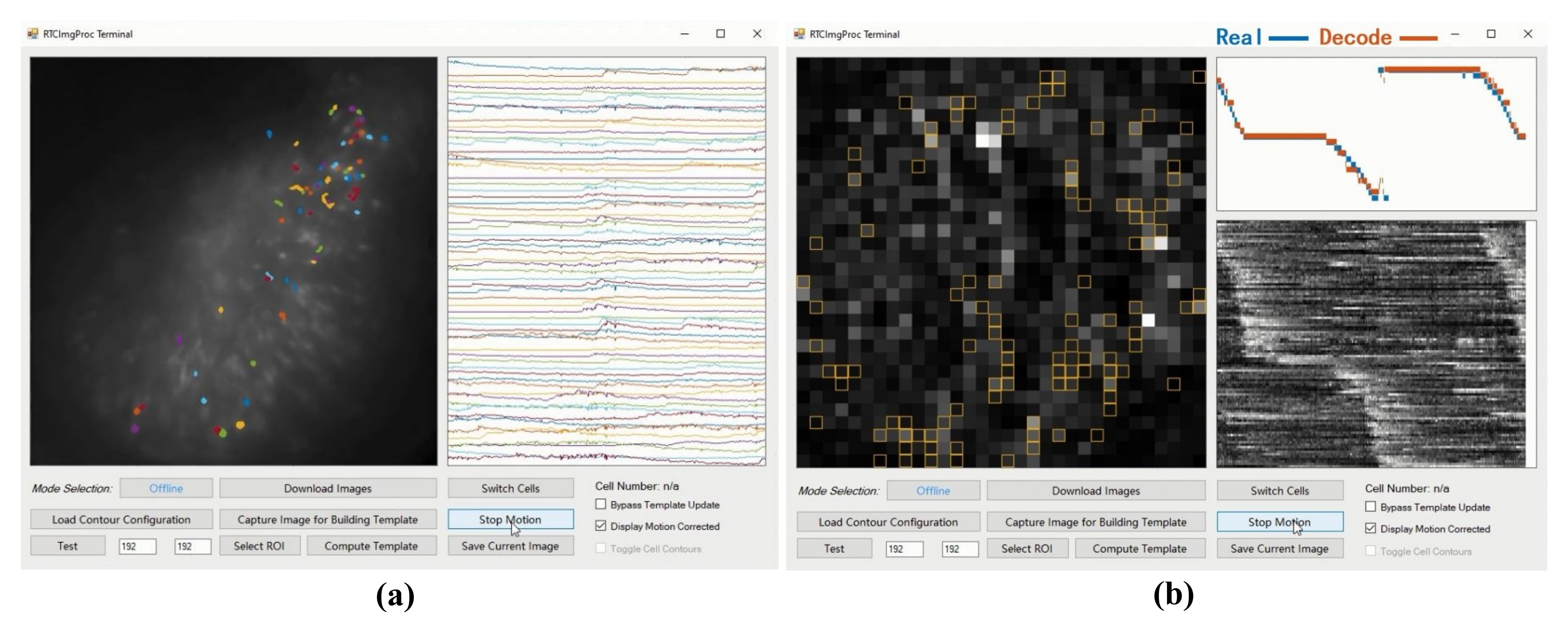}
\caption{Software interface for a demonstration on real-time (a) calcium image trace extraction and (b) calcium image position decoding.}
\label{fig:13}
\end{figure*}

We also developed graphical user interface software that runs on the host computer and interacts with the FPGA hardware. The user interface supports two application scenarios: 1) real-time calcium-image trace extraction; 2) real-time calcium image decoding. In the first scenario, the user interface is able to display received calcium images and a maximum of 63 extracted traces in real time. The contours of the selected cells can be superimposed on top of the displayed calcium images, as Fig. \ref{fig:13}(a) shows. In the second scenario shown in Fig. \ref{fig:13}(b), the user interface displays the tile-based trace values extracted at every frame with the real-time position decoding result. It also shows superimposed hollow contours and a flow of activity heat map for selected place tiles, which are identified offline \cite{Chen2022_eLife}.

\section{Evaluation} \label{sec:6}

\subsection{Performance Summary}

Our work achieves comparable performance against other state-of-the-arts targeting real-time calcium image processing for closed-loop applications. Table \ref{tab:6} summarizes a comparison among these works. \cite{Chen2022_eLife} and \cite{Liu2021} focus on real-time behavior decoding, whereas \cite{Zhang2018} and \cite{Taniguchi2021} attempt to address the real-time cell activity pattern detection problem. Compared with others, our work highlights the usage of hardware platform of the FPGA, and for the first time reduces the calcium image decoding latency to less than 1 ms, which is regarded as the biological spike-timing precision and critical timing factor for kilo-frame-rate voltage imaging \cite{Kazemipour2019}. Besides, our work first introduces the use of the SNN for the calcium image decoding, which introduces a new application area for the neuromorphic algorithms. Though the proposed rate-based SNN still cannot match the CNN model on the latency and thus the power consumption, it has shown the potential to become a low hardware cost counterpart aside from the CNN. Finally, our implementation not only supports closed-loop feedback, but it also provides an end-to-end tool flow with dedicated hardware and software interfaces for neuroscientists who rely on closed-loop experiments to advance brain research.

\begin{table*}
  \caption{Comparison with state-of-the-art calcium image processing and decoding implementations.}
  \label{tab:6}
  \centering
  \begin{tabular}{|l|c|c|c|c|}
    \hline
    & This Work & OBCIs \cite{Liu2021} & \cite{Zhang2018} & Carignan \cite{Taniguchi2021}\\
    \hline
    Calcium Imaging & One-photon & One-photon & Two-photon & One-photon\\
    \hline
    Spatial Resolution & 512×512 & 752×480 & 512×512 & 480×480\\
    \hline
    Frame Rate & 22.8-60 fps & 10-30 fps & 30 fps & 10 fps\\
    \hline
    Decode/Detect Task & Decode Position & Lever-pressing & Event Detection & Cell Detection\\
    \hline
    Decode/Detect Method & CNN/SNN/ANN & SVM & Threshold & CNN+LSTM\\
    \hline
    Contour Based Method & Yes & Yes & No & Yes\\
    \hline
    Num. of Cells/Tiles & Up to 1024 & 10 & 3 & 345-621\\
    \hline
    Decode/Detect Accuracy\textcolor{blue}{*} & 91.9\% (\textit{Hit-3}) & 81.04\% & 90\% & 89.4\%\\
    \hline
    Hardware Platform & Ultra96 FPGA & CPU & CPU, GPU & CPU\\
    \hline
    Processing Latency & $<$1 ms & 2.417 ms & 7.01 ms & 60.6 ms\\
    \hline
    End-to-End Solution & Yes & No & Yes & No\\
    \hline
    Closed-Loop Support & Yes & No & Yes & Yes\\
    \hline
    \multicolumn{5}{l}{*The accuracy evaluation metric is not the same across different works.}
  \end{tabular}
\end{table*}

\subsection{Comparison to State-of-the-Arts}

As miniaturized calcium imaging devices are gaining momentum as critical intervention tools for brain research, we have seen increased interests and related works focusing on calcium image analysis and processing. Based on targeted applications, these works can be mainly classified into 1) calcium imaging based neural signal extraction and 2) calcium image decoding.

Calcium imaging based neural signal extraction combines stabilization, enhancement, and spike inference to extract useful neural signal information from recorded calcium images. \cite{Pnevmatikakis2017} proposed a non-rigid motion correction method that is able to effectively compensate for non-ideal motion artifacts at subregions in the calcium image. \cite{Pnevmatikakis2014} and \cite{Pnevmatikakis2016} introduced the method for identifying cells from calcium images based on the CNMF approach. \cite{Pnevmatikakis2017} came up with an extension of the CNMF approach to address the background contamination for one-photon calcium images. \cite{Friedrich2016, Friedrich2017} and \cite{Pnevmatikakis2013} concentrated on spike inference from the calcium imaging data based on fast deconvolution and Bayesian methods. \cite{Giovannucci2017} with its follow-up work \cite{Giovannucci2019} unified previous efforts and established a complete calcium image analysis flow, which is open sourced with Python \cite{CaImAn-Python} and Matlab \cite{CaImAn-Matlab} releases. MIN1PIPE \cite{Lu2018} is another popular calcium image analysis pipeline which combines the LSTM inference for the cell identification and trace extraction. BSSE \cite{Son2019} offers a generic calcium image analysis tool for not only brain imaging but also other tissue studies. \cite{Stringer2019} brought Z-dimension motion registration and visualization to attention for improvement on calcium image analysis. \cite{Levin-Schwartz2017} proposed an independent component analysis based method to identify cells and extract neural signals from calcium images. Although these methods have been successful in extracting neural signal information from noisy calcium images, they are usually hard to be realized in real time. \cite{Giovannucci2017} and \cite{Giovannucci2019} achieve real-time calcium image analysis speed with optimized computation processes, but it's still hard to realize short processing latency for closed-loop applications. \cite{Taniguchi2021} proposed a closed-loop all-optical feedback stimulation framework based on threshold crossing detection of neural fluorescence signals, and it's dedicated for two-photon calcium imaging on head-fixing mice.

Some works look beyond neural activity and pursue the idea of calcium image decoding. \cite{Li2019} combined a deep residual neural network with the k-Means method to decode the forelimb reach activity from averaged calcium image from mice. \cite{Rubin2019} applied the Laplacian Eigenmaps to reduce dimension of neural activity extracted from calcium images and decode behavior states of mice on a linear track. \cite{Etter2020} studied the relationship between behavioral states and neuronal activity based on the Bayesian classifier. \cite{Wang2019} leveraged the MIN1PIPE \cite{Lu2018} for preprocessing, the principal component analysis for dimension reduction, and a linear SVM to decode the lever pressing movement of mice. \cite{Chen2018} benchmarked 5 different classifiers on a high/low velocity decoding task. Some other works look into tackling the real-time calcium image decoding challenge. \cite{Tu2020} showed that calcium traces without spike inference can be directly used for real-time position decoding by supervised and unsupervised methods. \cite{Lee2017} introduced incremental linear discriminant analysis method for real-time neural decoding with online adaptation capability. \cite{Liu2021} and \cite{Lee2019} developed real-time calcium image decoding pipelines based on the SVM method.

Compared to previous works, this work provides an end-to-end real-time calcium image processing and decoding solution for closed-loop feedback applications. Combining the customized algorithm pipeline, the accelerator design and optimization, and the hardware implementation, it achieves deterministic and sub-ms processing latency for closed-loop calcium image processing on the low-cost FPGA platform. It also evaluates and analyzes the tradeoff on accuracy, performance and cost among different neural network based methods for the calcium image decoding task.

\section{Conclusion} \label{sec:7}

This paper introduces an end-to-end FPGA-based prototype for real-time calcium image processing and decoding. With the support of dedicated hardware interface, it can perform a series of calcium image processing including the motion correction, enhancement, trace extraction and decoding in a frame-based fashion with sub-ms and deterministic processing latency. It empowers neuroscientists working with the in-vivo calcium imaging to perform efficient and precise closed-loop feedback stimulation for a wide range of neuroscientific experiments, and has potential to contribute to future brain research.

\section*{Acknowledgment}

The authors thank Carsten Hoffmann from AMD for donation of the Ultra96 board, Jeff Johnson for his advice on developing the Ethernet interface, and Dr. Changliang Guo for his support on the miniscope sensor and the DAQ board. This work was supported by NSF NeuroNex DBI-1707408.

\begin{IEEEbiography}[{\includegraphics[width=1in,height=1.25in,clip,keepaspectratio]{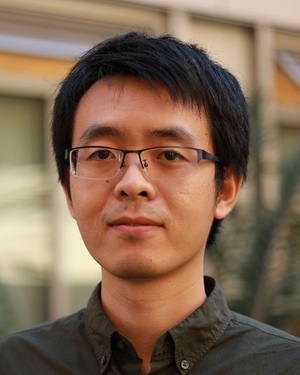}}]{Zhe Chen} (M'16) received the B.S. degree in electrical engineering from Tsinghua University in 2011, and the Ph.D. degree in solid-state circuits and microelectronics from the Institute of Semiconductors, Chinese Academy of Sciences in 2016. He started postdoctoral research at the University of California, Los Angeles in 2016 and became an Assistant Project Scientist in 2021. His research focused on acceleration of neural signal and image processing for closed-loop feedback applications, and his research interests include high performance and energy-efficient computing for machine vision. He won the Best Paper Award at the International Symposium on Low Power Electronics and Design in 2018, and received the UCLA Chancellor's Award for Postdoctoral Research in 2019.
\end{IEEEbiography}

\begin{IEEEbiography}[{\includegraphics[width=1in,height=1.25in,clip,keepaspectratio]{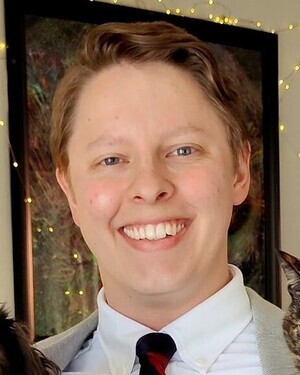}}]{Garrett James Blair} received the B.S. degree in cognitive science, the B.A. degree in Spanish literature from the University of California, Irvine in 2015, and the Ph.D. degree in Psychology from the University of California, Los Angeles under supervision of Prof. Hugh T. Blair in 2021. His research interest seeks to understand how the brain synthesizes and organizes information to learn and navigate the world. He is currently a postdoctoral fellow at New York University in Dr. André Fenton's lab evaluating the neural basis of cognitive control. Dr. Blair has been the recipient of the UCLA Alumni fellowship in 2015, the UCLA-Peking University joint research scholarship in 2016, and has published across diverse research areas.
\end{IEEEbiography}

\begin{IEEEbiography}[{\includegraphics[width=1in,height=1.25in,clip,keepaspectratio]{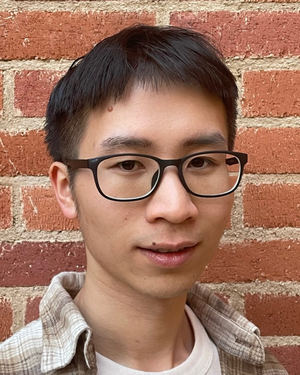}}] {Chengdi Cao} received the B.S. degree in electrical engineering from the Tsinghua University, Beijing, China. Since 2021, he has been working towards the Ph.D. degree in computer science under the guidance of Prof. Jason Cong and Prof. Cho-Jui Hsieh at the University of California, Los Angeles. His current research interests include efficient machine learning and high-level synthesis.
\end{IEEEbiography}

\begin{IEEEbiography}[{\includegraphics[width=1in,height=1.25in,clip,keepaspectratio]{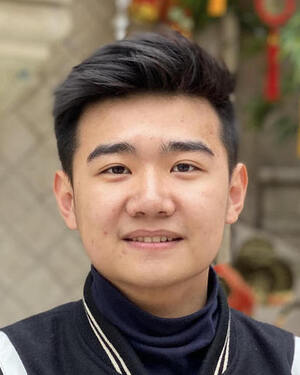}}] {Jim Zhou} received the B.S. degree in computer science from the University of California Los Angeles, Los Angeles, California and he is currently working towards the M.S. degree in computer science at Yale University, New Haven, Connecticut. He is the recipient of the Third Place Award at the IEEE/ACM International Symposium on Microarchitecture (MICRO) Student Research Competition in 2021. His research interests include spiking neural networks, neuromorphic hardware and the FPGA.
\end{IEEEbiography}

\begin{IEEEbiography}[{\includegraphics[width=1in,height=1.25in,clip,keepaspectratio]{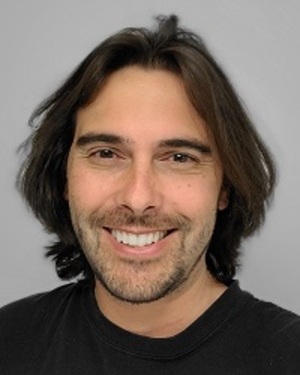}}] {Daniel Aharoni} received his B.S and Ph.D degrees in physics from the University of California, Los Angeles in 2006 and 2013, respectively. He is currently an Assistant Professor in the Department of Neurology at the University of California, Los Angeles. Dr. Aharoni's research interests are at the intersection of engineering, neuroscience, and physics. Specifically, his work focuses on applying tool development methodologies from engineering and physics to address current challenges in neuroscience and medicine. Dr. Aharoni led the development of the Miniscope system, an open-source microscopy platform, which is currently being used in over 450 laboratories with an active and growing user base, making it one of the most successful open-source neuroscience tools to date.
\end{IEEEbiography}

\begin{IEEEbiography}[{\includegraphics[width=1in,height=1.25in,clip,keepaspectratio]{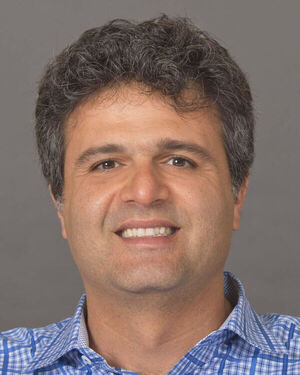}}] {Peyman Golshani} obtained his M.D. and Ph.D. degrees from the University of California, Irvine and the University of California, Davis in 2002. He has been a faculty at University of California, Los Angeles since 2006 and he is currently a Professor in the Departments of Neurology and Psychiatry. His laboratory develops techniques for recording large-scale activity patterns in freely behaving animals and studies how these activity patterns drive attention, working memory and long-term memories. 
\end{IEEEbiography}

\begin{IEEEbiography}[{\includegraphics[width=1in,height=1.25in,clip,keepaspectratio]{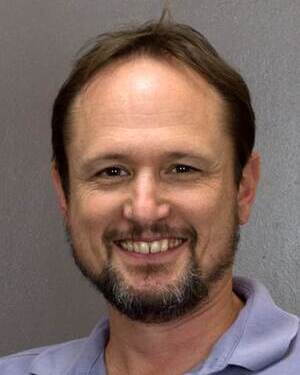}}] {Hugh Tad Blair} received the B.S. degree in communications from the University of Texas at Austin in 1989, the M.S. degree in computer science from Northwestern University in 1993, and the Ph.D. degree in behavioral neuroscience from Yale University in 1999. He is currently a Professor at the University of California, Los Angeles and the Area Chair for the Behavioral Neuroscience in the Department of Psychology. His laboratory integrates experimental and theoretical approaches to investigate the neurobiological basis for learning, memory, and decision making. He is also active in the development of new technologies for neuron-resolution calcium imaging and electrophysiology.
\end{IEEEbiography}

\begin{IEEEbiography}[{\includegraphics[width=1in,height=1.25in,clip,keepaspectratio]{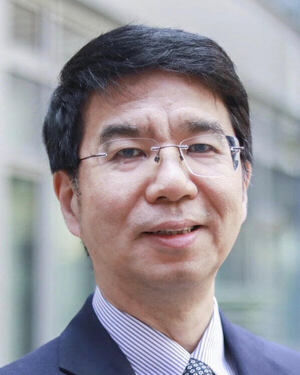}}] {Jason Cong} (F'00) received the B.S. degree from Peking University in 1985, and M.S. and Ph.D. degrees from the University of Illinois at Urbana-Champaign in 1987 and 1990, respectively, all in computer science. He is currently the Volgenau Chair for Engineering Excellence Professor at the UCLA Computer Science Department (and a former Department Chair), with joint appointment from the Electrical and Computer Engineering Department. He is the Director of the Center for Domain-Specific Computing (CDSC) and the Director of the VLSI Architecture, Synthesis, and Technology (VAST) Laboratory.
  
Dr. Cong's research interests include novel architectures and compilation for customizable computing, synthesis of VLSI circuits and systems, and quantum computing. He has over 500 publications in these areas, including 17 Best Paper Awards, and 4 papers in the FPGA and Reconfigurable Computing Hall of Fame. He and his former students co-founded AutoESL, which developed the most widely used high-level synthesis tool for FPGAs (renamed to Vivado HLS and Vitis HLS after Xilinx's acquisition). He is a Member of the National Academy of Engineering, and a Fellow of ACM, IEEE, and the National Academy of Inventors. He is the recipient of the 2022 IEEE Robert Noyce Medal for fundamental contributions to electronic design automation and FPGA design methods.
\end{IEEEbiography}

\end{document}